\documentclass[aip,rsi, amsmath,amssymb, reprint, fontsize=12pt]{revtex4-1}
\pdfoutput=1
\usepackage{graphicx}
\usepackage{dcolumn}
\usepackage{bm}

\usepackage{pgfplots}
\usepackage{pgfplotstable}
\usepgfplotslibrary{external} 
\usepgfplotslibrary{fillbetween}
\usepgfplotslibrary{colorbrewer}
\pgfplotsset{compat=1.6}
\usepackage{lineno}
\bibpunct{[}{]}{;}{n}{}{,~}

\usetikzlibrary{external}
\usepackage{subfigure}
\usepackage{cleveref}

\pgfmathsetmacro{\Tfreeze}{364}
\pgfmathsetmacro{\Tjump}{442.335666891}
\pgfmathsetmacro{\tfreeze}{0.483233608052}
\pgfmathsetmacro{\tjump}{0.535762818744}
\pgfmathsetmacro{\Tc}{537}

\definecolor{android_blue}{RGB}{51,181,229}
\definecolor{android_dark_blue}{RGB}{0,153,204}
\definecolor{android_pink}{RGB}{170,102,204}
\definecolor{android_purple}{RGB}{156,39,176}
\definecolor{android_dark_pink}{RGB}{153,51,204}
\definecolor{android_green}{RGB}{153,204,0}
\definecolor{android_dark_green}{RGB}{102,153,0}
\definecolor{android_orange}{RGB}{255,152,0}
\definecolor{android_dark_orange}{RGB}{255,152,0}
\definecolor{android_red}{RGB}{255,68,68}
\definecolor{android_dark_red}{RGB}{204,0,0}
\definecolor{android_pink}{RGB}{156,39,176}
\definecolor{android_grey}{RGB}{158,158,158}

\definecolor{olivia_red}{RGB}{215,25,28}
\definecolor{olivia_orange}{RGB}{253,174,97}
\definecolor{olivia_yellow}{RGB}{255,255,191}
\definecolor{olivia_lightblue}{RGB}{171,217,233}
\definecolor{olivia_blue}{RGB}{44,123,182}

\pgfplotsset{grid style={dashed,grey,opacity=0.5}}

\pgfplotscreateplotcyclelist{peak_temp}{
{color=android_dark_blue,line width=1.0pt,mark=x,mark size=2pt,mark options={line width=1.0pt},line join=round},
{color=android_red,line width=1.0pt,mark=o,mark size=2pt,mark options={line width=1.0pt},line join=round},
{color=android_dark_green,line width=0.5pt,mark=|,mark size=2pt,mark options={line width=0.75pt},line join=round},
{color=black,line width=0.75pt,mark size=2pt,mark=triangle,mark options={line width=0.75pt},line join=round},
{color=android_blue,line width=0.5pt,mark=square,mark size=2pt,mark options={line width=0.75pt},line join=round},
	  {color=android_pink,line width=0.5pt,mark=diamond,mark size=2pt,mark options={line width=0.75pt},line join=round},
	  {color=android_orange,line width=0.5pt,mark size=2pt,mark options={line width=0.75pt},line join=round}
	  }

\begin{document}
\pgfplotsset{colormap/RdBu-9}

\title{Improving the Signal-to-noise Ratio for Heat-Assisted Magnetic Recording by Optimizing a High/Low Tc bilayer structure}

\author{O. Muthsam}
 \email{olivia.muthsam@univie.ac.at}
\author{F. Slanovc}
\author{C. Vogler}
\author{D. Suess}
\affiliation{ 
University of Vienna, Physics of Functional Materials, Boltzmanngasse 5, 1090 Vienna, Austria
}%

\date{\today}
             
\begin{abstract}
We optimize the recording medium for heat-assisted magnetic recording by using a high/low $T_{\mathrm{c}}$ bilayer structure to reduce AC and DC noise. Compared to a former work, small Gilbert damping $\alpha=0.02$ is considered for the FePt like hard magnetic material. Atomistic simulations are performed for a cylindrical recording grain with diameter $d=5\,$nm and height $h=8\,$nm. Different soft magnetic material compositions are tested and the amount of hard and soft magnetic material is optimized. 
The results show that for a soft magnetic material with $\alpha_{\mathrm{SM}}=0.1$ and $J_{ij,\mathrm{SM}}=7.72\times 10^{-21}\,$J/link a composition with $50\%$ hard and $50\%$ soft magnetic material leads to the best results. Additionally, we analyse how much the areal density can be improved by using the optimized bilayer structure compared to the pure hard magnetic recording material. It turns out that the optimized bilayer design allows an areal density that is 1\,Tb/in$^2$ higher than that of the pure hard magnetic material while obtaining the same SNR.

\end{abstract}

\maketitle

\section{Introduction}

Heat-assisted magnetic recording (HAMR) \cite{kobayashi,mee,rottmayer,kryder,rausch_heat_2015,thermomagnetic,burns} is a promising recording technology to further increase the areal storage densities (ADs) of hard disk drives. Conventional state-of-the-art recording technologies are not able to overcome the so-called recording trilemma \cite{evans2}: Higher ADs require smaller grains. These grains need to have high uniaxial anisotropy to be thermally stable. However today's write heads are not able to produce fields that are strong enough to switch these high anisotropy grains. In the HAMR process a heat pulse is included in the recording process to locally heat the recording medium. This leads to a drop of the coercivity, making the high anisotropy recording medium writeable. The medium is then quickly cooled and the information reliably stored.\\
To reach high linear densities it is necessary to reduce AC and DC noise in recording media \cite{basic}. AC noise determines the distance between neighboring bits in bit-patterned \cite{bitpatterned1,bitpatterned2,yen_bit_2013} media or the transition between grains in granular media. DC noise restricts the maximum switching probability of grains away from the transition. It has been shown, that pure hard magnetic grains do not switch reliably \cite{fundamental} if bit-patterned media are considered whereas non-optimized exchange coupled bilayer structures \cite{suess,suessexchange,victora,wangexchange,coffey,suess1} of hard and soft magnetic material experience high AC noise \cite{areal}. A work to reduce noise in recording media by optimizing a high/low $T_{\mathrm{c}}$ bilayer structure (see Ref. \cite{noisehamr}) showed that an optimial bilayer structure consists of 80$\%$ hard magnetic and 20$\%$ soft magnetic material. However, in the former work the Gilbert damping was assumed to be $\alpha_{\mathrm{HM}}=0.1$ which is hard to achieve in a FePt like hard magnetic material in reality. In realistic hard magnetic recording materials, the damping constant is $\alpha=0.02$, according to the Advanced Storage Technology Consortium (ASTC) \cite{ASTC}.
Since it has been shown that the damping constant has a strong influence on the maximum switching probability and the down-track jitter, we follow the optimization approach and optimize a bilayer structure for the ASTC parameters. After the optimization, we study how the optimized material differs from that with $\alpha_{\mathrm{HM}}=0.1$.\\
Additionally, we investigate how much the areal storage density (AD) can be improved when using the optimized recording material instead of the pure hard magnetic one. This is done with the help of the signal-to-noise ratio (SNR), which gives the power of the signal over the power of the noise and is a good indicator for the quality of written bits.\\
The structure of this work is as follows: 
In Section II, the HAMR model and the material parameters are presented. In Section III, the results are shown and they are discussed in Section IV.

\section{HAMR model}
The optimization simulations are performed with the atomistic simulation program VAMPIRE \cite{evans} which solves the stochastic Landau-Lifshitz-Gilbert (LLG) equation. 
In the simulations, a cylindrical recording grain with a diameter $d=5\,$nm and a height $h=8\,$nm is used. It can be considered as one recording bit in bit-patterned media. A simple cubic crystal structure is used and only nearest neighbor interactions are considered. The effective lattice parameter $a$ and the exchange interaxtion $J_{ij}$ are adjusted in order to lead to the experimentally obtained saturation magnetization and Curie temperature. \cite{mryasov2005temperature,hovorka2012curie}. The write head is assumed to move with a velocity of $v=15\,$m/s. A continuous laser pulse is assumed with the Gaussian temperature profile

\begin{align}
T(x,y,t)= (T_{\mathrm{write}}-T_{\mathrm{min}})e^{-\frac{x^2+y^2}{2\sigma^2}} + T_{\mathrm{min}} \\
= T_{\mathrm{peak}}(y)\cdot e^{-\frac{x^2}{2\sigma^2}} + T_{\mathrm{min}}
\label{pulse}
\end{align}
with
\begin{align}
\sigma=\frac{\mathrm{FWHM}}{\sqrt{8\ln(2)}}.
\end{align}

The full width at half maximum (FWHM) is assumed to be 60\,nm. Both, the down-track position $x$ and the off-track position $y$ are variable in the simulations. 
The initial and final temperature is $T_{\mathrm{min}}=300\,$K. The applied field is modeled as a trapezoidal field with a write field duration of 0.57\,ns and a field rise and decay time of 0.1\,ns. The field is applied at an angle of 22\,deg with respect to the normal. The field strength is assumed to be +0.8\,T and -0.8\,T in $z$-direction. Initially, the magnetization of each grain points in $+z$-direction. The trapezoidal field tries to switch the magnetization of the grain from $+z$-direction to $-z$-direction. At the end of every simulation, it is evaluated if the bit has switched or not.\\

\subsection{Material parameters}

The material parameters for the hard magnetic material can be seen in \Cref{opt2tablematerialien}. For the soft magnetic material, the atomistic spin moment is assumed to be $\mu_{\mathrm{s}}=1.6\,\mu_{\mathrm{B}}$ which corresponds to a saturation polarization $J_{\mathrm{s}}=1.35\,$T. The uniaxial anisotropy constant $k_{\mathrm{u,SM}}$ in the soft magnetic layer is initially set to 0 but later varied. The Gilbert damping $\alpha_{\mathrm{SM}}$ and the exchange interaction $J_{ij,SM}$ within the soft magnetic material are varied. Experimentally, it is possible to increase the damping constant by doping the soft magnetic material with transition metals like Gd or Os \cite{doping,ingvarsson_tunable_2004,fassbender_structural_2006,bailey_control_2001,rantschler_effect_2007}. Thus, also enhanced damping constants $\alpha_{\mathrm{SM}}$ larger than $0.02$ are considered in the simulations.\\

\begin{center}
\begin{table*}
\centering
\begin{tabular}{|>{\centering}m{3.0cm}|>{\centering}m{3.0cm}|>{\centering}m{4.1cm}|>{\centering}m{2.5cm}|c|}
\hline
Curie temp. $T_{\mathrm{\textbf{C}}}$ [K] & Damping $\alpha$& Uniaxial anisotropy. $k_{u}$ [J/link]& $J_{ij}$ [J/link]& $\mu_{\mathrm{s}}$  [$\mu_{\mathrm{B}}$] \\
    \hline
	693.5 & 0.02&$9.124\times10^{-23}$  & $6.72 \times 10^{-21}$ &1.6\\
    \hline
 \end{tabular}
\caption{Material parameters of a FePt like hard magnetic granular recording medium. }
\label{opt2tablematerialien}
\end{table*}
\end{center}

\section{Results}

\subsection{Hard magentic grain}
\begin{figure}
\centering
\includegraphics[width=1.0\linewidth]{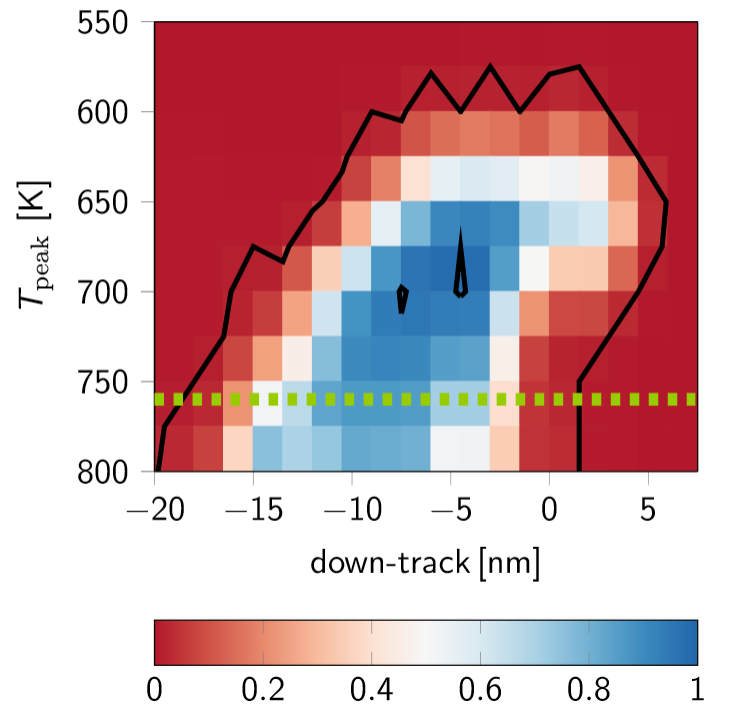}
  \caption{Switching probability phase diagram of a pure FePt like hard magnetic grain. The contour lines indicate the transition between areas with switching probability less than 1$\%$ (red) and areas with switching probability higher than 99.2$\%$ (blue). The dashed lines mark the switching probability curves of \Cref{opt2feptdowntrack}. }
  \label{opt2feptphase}
\end{figure}

First, a switching probability phase diagram for the pure hard magnetic material is computed where the switching probability is depending on the down-track position $x$ and the off-track position $y$. With \cref{pulse} each off-track position $y$ can be transformed into an unique peak temperature $T_{\mathrm{peak}}$, if the write temperature $T_{\mathrm{write}}$ is fixed, and vice versa. Thus, the switching probability in \Cref{opt2feptphase} is shown as a function of the down-track position $x$ and the peak temperature $T_{\mathrm{peak}}$ that corresponds to $y$. The resolution of the phase diagram in down-track direction is $\Delta x=1.5\,$nm and that in temperature direction is $\Delta T_{\mathrm{peak}}=25\,$K. In each phase point, 128 trajectories are simulated with a simulation length of $1.5\,$ns. Thus, the phase diagram contains more than 30.000 switching trajectories. From the phase diagram it can be seen that the pure hard magnetic grain shows only two small areas with switching probability larger than $99.2\%$. This threshold is used, since 128 simulations per phase point are performed and a switching probability of 100$\%$ corresponds to a number of successfully switched trajectories larger than $1-1/128=0.992$.

\begin{figure}
\centering
\includegraphics[width=0.8\linewidth]{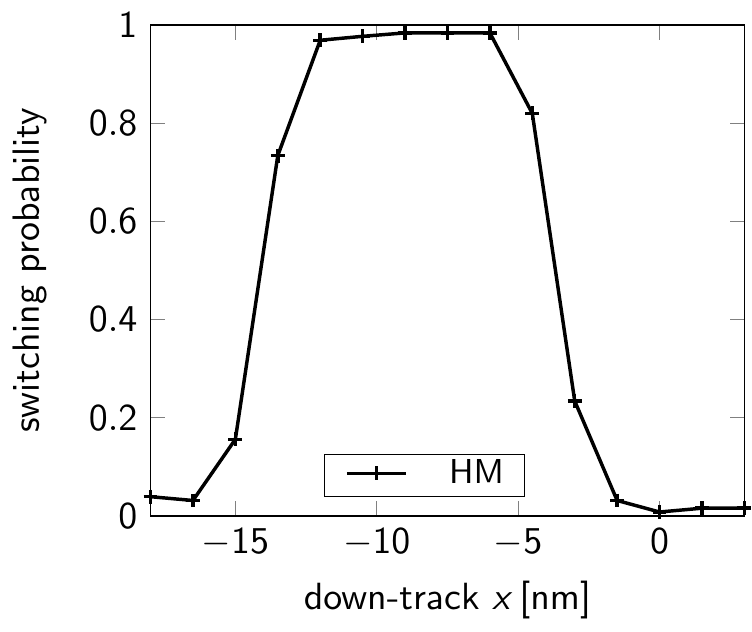}
  \caption{ Down-track switching probability curve $P(x)$ at a peak temperature $T_{\mathrm{peak}}=760$\,K for a pure hard magnetic grain.}
  \label{opt2feptdowntrack}
\end{figure}

To determine the down-track jitter $\sigma$, a down-track switching probability curve $P(x)$ for $-20\,$nm $\le x \le 6\,$nm at a fixed temperature $T_{\mathrm{peak}}=760\,$K is determined for pure hard magnetic material (see \Cref{opt2feptdowntrack}). The switching probability curve is fitted with a Gaussian cumulative function

\begin{align}
\Phi_{\mu,\sigma^2}=\frac{1}{2} (1 + \mathrm{erf}(\frac{x-\mu}{\sqrt{2\sigma^2}}))\cdot P
\label{distribution}
\end{align}
with
\begin{align}
\mathrm{erf}(x)=\frac{2}{\sqrt{\pi}} \int_0^x e^{-\tau^2} d\tau,
\label{error}
\end{align}

where the standard deviation $\sigma$, the mean value $\mu$ and the mean maximum switching probability $P \in [0,1]$ are the fitting parameters. The standard deviation $\sigma$ determines the steepness of the transition function and is a measure for the transition jitter. In the further course it will be called $\sigma_{\mathrm{down}}.$ The fitting parameter $P$ is a measure for the average switching probability for sufficiently high temperatures. The resulting fitting parameters of the hard magnetic material can be seen in \Cref{opt2tableratios}. Note, that the calculated jitter values only consider the down-track contribution of the write jitter. The so-called $a-$parameter is given by

\begin{align}
    a =  \sqrt{\sigma_{\mathrm{down}}^2+\sigma_{\mathrm{g}}^2}
\end{align}

where $\sigma_{\mathrm{g}}$ is a grain-size-dependent jitter contribution \cite{wang_transition_2009}. The write jitter can then be calculated by

\begin{align}
    \sigma_{\mathrm{write}} \approx a \sqrt{\frac{S}{W}}
\end{align}

where $W$ is the reader width and $S=D+B$ is the grain diameter, i.e. the sum of the particle size $D$ and the nonmagnetic boundary $B$ \cite{varvaro_ultra-high-density_2016,slanovc}. 

\subsection{Media Optimization}
\begin{figure}
\centering
\includegraphics[width=1.0\linewidth]{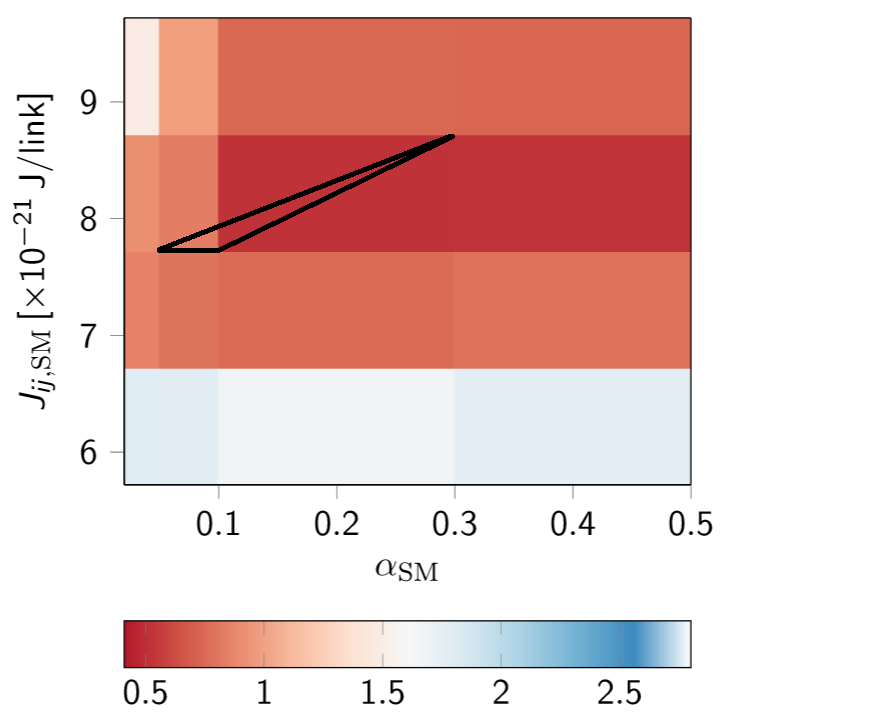}
  \caption{Down-track jitter $\sigma_{\mathrm{down}}$ as a function of the damping constant and the exchange interaction. The contour line indicates the transition between areas with down-track jitter larger than 0.5\,nm (light red, blue) and areas with down-track jitter smaller than 0.5\,nm (dark red).  }
  \label{opt2alphavsJij}
\end{figure}

To find the best soft magnetic material composition, down-track switching probability curves $P(x)$ similar to \Cref{opt2feptdowntrack} are computed for 50/50 bilayer structures with different damping constants $\alpha_{\mathrm{SM}}$ and different exchange interactions $J_{ij,\mathrm{SM}}$. The range in which the parameters are varied can be seen in \Cref{opt2tablevalues}. Note, that $P(x)$ is computed at different peak temperatures for the different exchange interactions, since there holds

\begin{align}
J_{ij}= \frac{3 k_{\mathrm{B}} T_{\mathrm{C}}}{\epsilon z},
\end{align}

where $k_{\mathrm{B}}$ is the Boltzmann constant, z is the number of nearest neighbors and $\epsilon$ is a correction factor from the mean-field expression which is approximately 0.86 \cite{evans}. The temperature at which $P(x)$ is calculated is chosen to be $T_{\mathrm{C}}+60\,$K. The down-track switching probability curves are then fitted with \cref{distribution}. 
The down-track jitter parameters as a function of the damping constant and the exchange interaction can be see in \Cref{opt2alphavsJij}. The maximum switching probability is 1 for $\alpha \ge 0.1$.

\begin{center}
\begin{table*}
\centering
\begin{tabular}{|l|c|c|c|c|}\hline
Parameter & min. value & max.value\\
\hline
$\alpha_{\mathrm{SM}}$&0.02  &0.5\\
$J_{ij,\mathrm{SM}}$\,[J/link]	&$5.72\times 10^{-21}$ &$9.72\times 10^{-21}$ \\
$k_{u,\mathrm{SM}}$\,[J/link]	&0& $1/2 k_{\mathrm{u,HM}}=4.562\times10^{-23}$\\
  \hline
 \end{tabular}
\caption{Range in which the different soft magnetic material parameters are varied.}
\label{opt2tablevalues}
\end{table*}
\end{center}

\begin{center}
\begin{table*}
\centering
\begin{tabular}{|l|c|c|c|c|}\hline
$k_{\mathrm{u,SM}}\,\times10^{-23}$\,[J/link] & $\sigma_{\mathrm{down}}$\,[nm] &$P$\,\\
\hline
$0$&0.41&1.0\\
  $0.562$&0.919&1.0\\
   $1.8428\,\, [=1/5\,k_{\mathrm{u,HM}}]$&1.04&1.0\\
   $3.124$&0.898&1.0\\
   $4.562\,\, [=1/2\,k_{\mathrm{u,HM}}]$&1.01&1.0\\
   
   \hline
 \end{tabular}
\caption{Resulting down-track jitter parameters and mean maximum switching probability values for soft magnetic materials with different uniaxial anisotropy constants $k_{\mathrm{u,SM}}$.}
\label{opt2tablek1}
\end{table*}
\end{center}

\begin{center}
\begin{table*}
\centering
\begin{tabular}{|>{\centering}m{3.0cm}|>{\centering}m{4.1cm}|>{\centering}m{2.5cm}|c|}
\hline
 Damping $\alpha_{\mathrm{SM}}$& Uniaxial anisotropy. $k_{u}$ [J/link]& $J_{ij}$ [J/link]& $\mu_{\mathrm{s}}$  [$\mu_{\mathrm{B}}$] \\
    \hline
	0.1&$0$  & $7.72 \times 10^{-21}$ &1.6\\
    \hline
 \end{tabular}
\caption{Resulting material parameters for the optimal soft magnetic material composition. }
\label{opt2tablesoft}
\end{table*}
\end{center}

\begin{center}
\begin{table*}
\centering
\begin{tabular}{|l|c|c|c|c|}\hline
 HM/SM & $\sigma_{\mathrm{down}}$\,[nm] &$P$\,\\
\hline
HM&0.974&0.95\\
  	90/10&1.06&0.969\\
	80/20& 0.813 & 0.998 \\
   70/30 & 0.6 & 0.988\\
   60/40 & 0.8& 0.999\\
   50/50 & 0.41 & 1.0\\
  \hline
 \end{tabular}
\caption{Resulting down-track jitter parameters and mean maximum switching probability values for hard magnetic material and three different hard/soft bilayer structures with different damping constants in the soft magnetic material.}
\label{opt2tableratios}
\end{table*}
\end{center}

\begin{figure}
\centering
\includegraphics[width=1.0\linewidth]{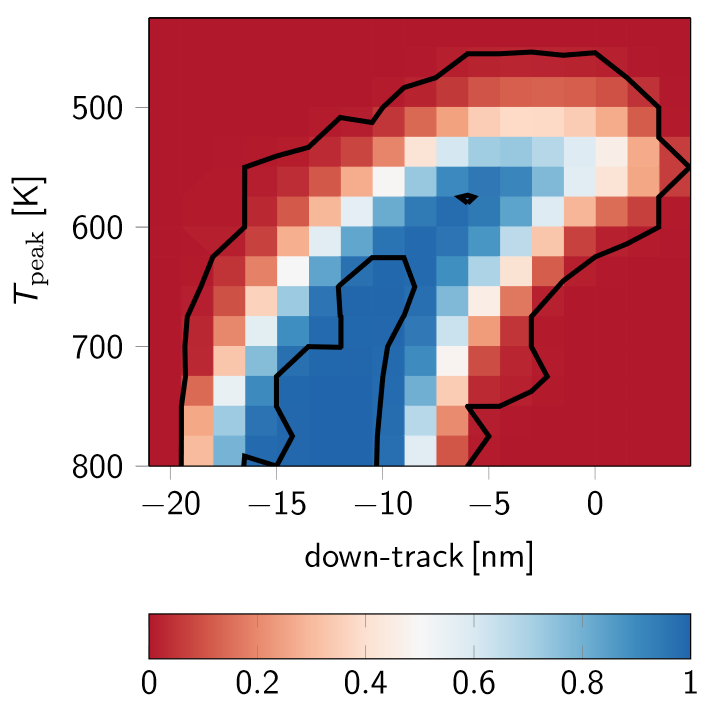}
  \caption{Switching probability phase diagram of recording grain consisting of a composition of 50$\%$ hard magnetic material and 50$\%$ soft magnetic material with $k_{\mathrm{u,SM}}=0\,$J/link and $J_{ij,\mathrm{SM}}=7.72\times 10^{-21}\,$J/link. 
The contour lines indicate the transition between areas with switching probability less than 1$\%$ (red) and areas with switching probability higher than 99.2$\%$ (blue).}
  \label{opt2hmsmphase}
\end{figure}

From the simulations it can be seen that a Gilbert damping $\alpha_{\mathrm{SM}}=0.1$ together with $J_{ij,\mathrm{SM}}=7.72\times 10^{-21}\,$J/link leads to the best results with the smallest down-track jitter $\sigma_{\mathrm{down}}=0.41\,$nm and a switching proability $P=1$. \\
The last soft magnetic parameter that is varied, is the uniaxial anisotropy $k_{\mathrm{u,SM}}$. It is known that the smallest coercive field in an exchange spring medium can be achieved if $K_{\mathrm{SM}}=1/5K_{\mathrm{HM}}$ \cite{hagedorn_analysis_1970,suess_multilayer_2006}. Here 

\begin{align}
    K_{\mathrm{i}}=\frac{n_{\mathrm{at}}k_{\mathrm{u,i}}}{a^3}\,\, i\in \{\mathrm{SM},\mathrm{HM}\}
\end{align}

are the macroscopic anisotropy constants in J/m$^3$ with the unit cell size $a=0.24\,$nm and the number of atoms $n_{\mathrm{at}}$ per unit cell. $k_{\mathrm{u,SM}}$ is varied between 0 and $1/2 k_{\mathrm{u,HM}}=4.562\times10^{-23}\,$J/link. The damping constant is $\alpha_{\mathrm{SM}}=0.1$. The resulting fitting parameters are summarized in \Cref{opt2tablek1}. It can be seen that the switching probability is one for all varied $k_{\mathrm{u,SM}}$. However, the down-track jitter increases for higher $k_{\mathrm{u,SM}}$. Since for $k_{\mathrm{u,SM}}=0\,$J/link the jitter is the smallest, this value is chosen for the optimal material composition.\\
In conclusion, the material parameters of the optimized soft magnetic material composition can be seen in \Cref{opt2tablesoft}.\\
Next, simulations for different ratios of hard and soft magnetic material are performed. Down-track switching probability curves $P(x)$ are computed for different ratios at $T_{\mathrm{peak}}=780\,$K and the down-track jitter and the mean maximum switching probability are determined. The results are listed in \Cref{opt2tableratios}.\\
It can be seen that a structure with $50\%$ hard magnetic and $50\%$ soft magnetic materials leads to the smallest jitter and the highest switching probability. This result differs from the optimized material composition in Ref. \cite{noisehamr}, where the optimal composition consists of $80\%$ hard magnetic and $20\%$ soft magnetic materials.
In \Cref{opt2hmsmphase}, a switching probability phase diagram of the optimized bilayer structure with $50\%$ hard and $50\%$ soft magnetic material can be seen.

It is visible that the switching probability of the structure is larger than $99.2\%$ for a bigger area of down-track positions and peak temperatures. This shows the reduction of DC noise in the optimized structure.\\

\subsection{Areal Density}
To analyse the possible increase of areal density by using the optimized bilayer structure instead of the pure hard magnetic recording medium, the signal-to-noise ratio is calculated. With the help of an analytical model of a phase diagram developed by Slanovc \textit{et al} \cite{slanovc} it is possible to calculate a switching probability phase diagram from eight input parameters. The input parameters are $P_{\mathrm{max}}$, $\sigma_{\mathrm{down}},$ the off-track jitter $\sigma_{\mathrm{off}},$ the transition curvature, the bit length, the half maximum temperature and the position of the phase diagram in $T_{\mathrm{peak}}$ direction and the position of the phase diagram in down-track direction. The $\sigma_{\mathrm{down}}$ and $P_{\mathrm{max}}$ values are those resulting from the simulations for pure hard magnetic material and the optimized bilayer structure. All other model input parameters are obtained by a least square fit from a switching probability phase diagram computed with a coarse-grained LLB model \cite{vogler_landau-lifshitz-bloch_2014}. The phase diagram is mapped onto a granular recording medium where the switching probability of the grain corresponds to its position. The writing process is repeated for 50 different randomly initialized granular media. The SNR is then computed from the read-back process with the help of a SNR calculator provided by SEAGATE \cite{hernandez_using_2017}. \\
The SNR is analysed for areal densities of 2 to 5\,Tb/in$^2$. 
%
For the bitsize ($bs$) at a certain areal density, there are different track width and bit length combinations ($t,b$) that yield 

\begin{align}
    bs=t\cdot b.
\end{align}

\begin{figure}
\centering
\subfigure{\includegraphics[width=1.0\linewidth]{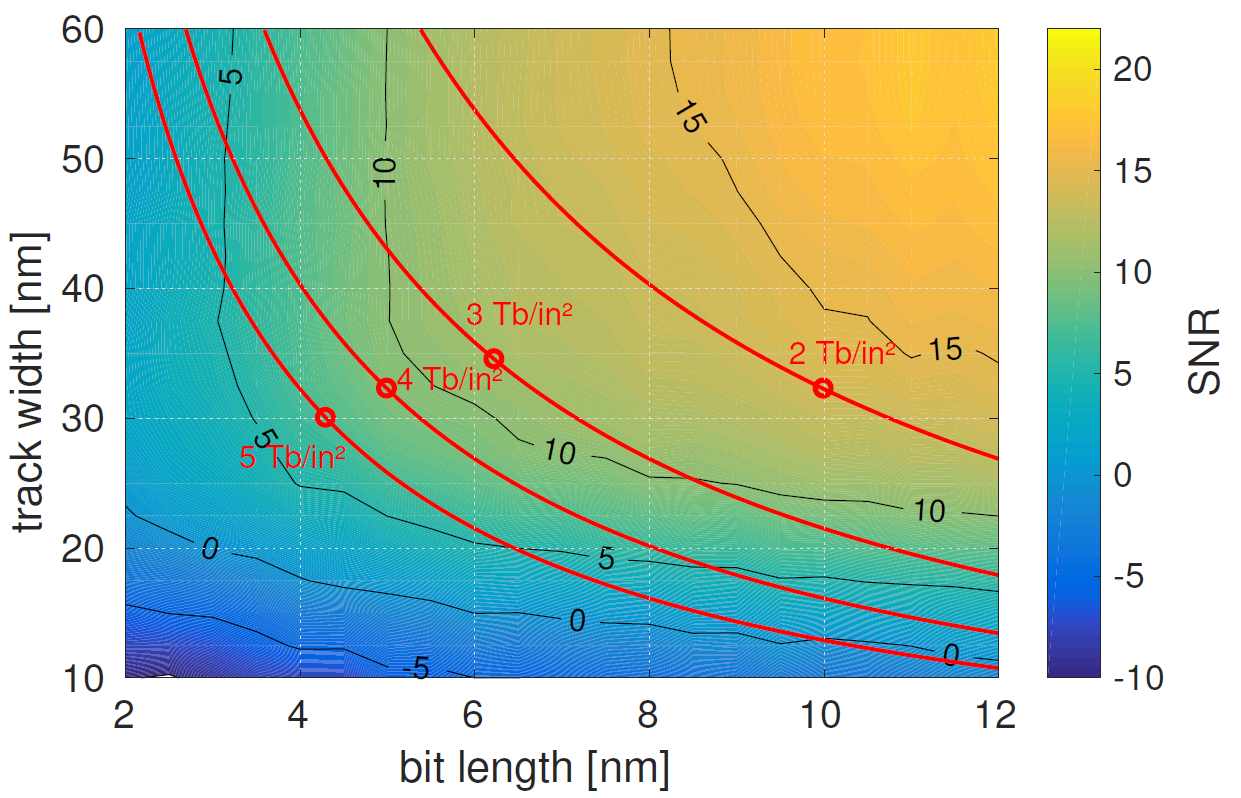}}
\subfigure{\includegraphics[width=1.0\linewidth]{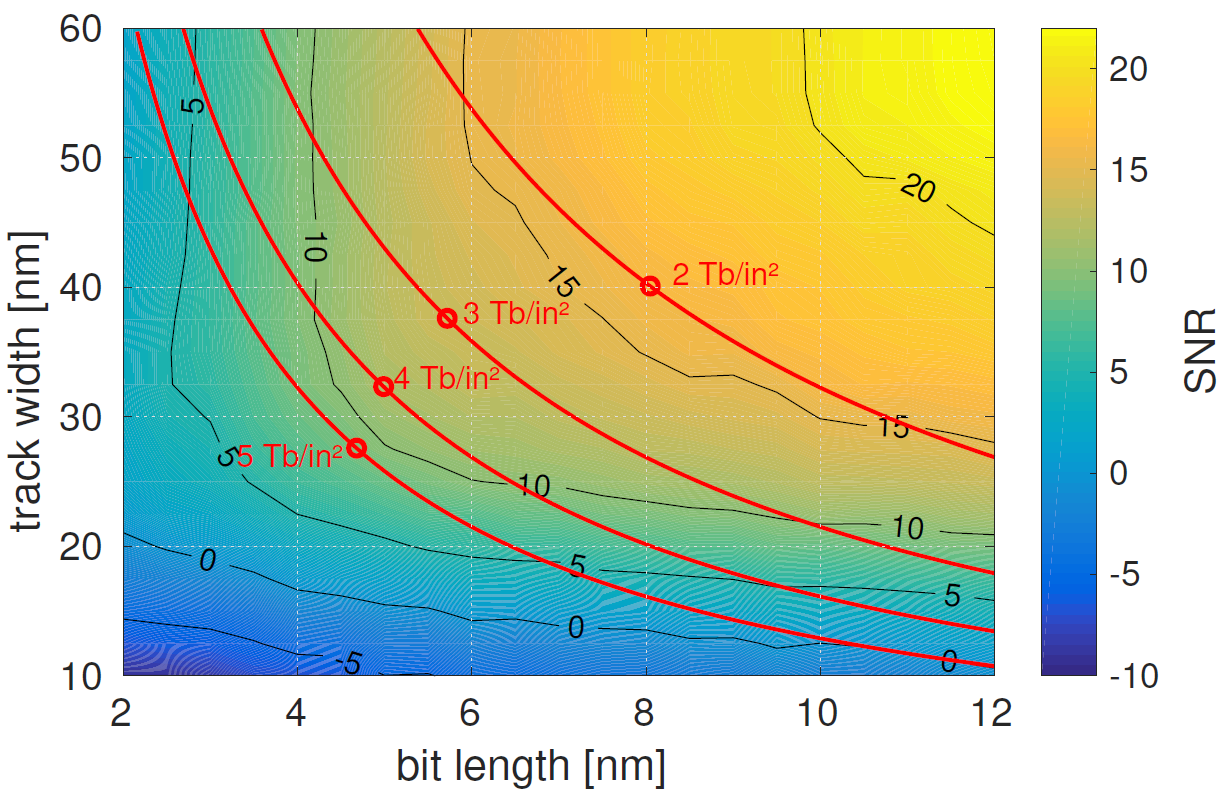}}
\caption{Signal-to-noise ratio (in dB) as a function of the bit length and the track width for (a) pure hard magnetic material and (b) the optimized hard/soft bilayer structure. The red lines indicate the bit length $-$ track width combinations that yield 2, 3, 4 and 5\,Tb/in$^2$ areal density. The dots indicate the combination at which the SNR is maximal.}
\label{opt2SNRphase}
\end{figure}

\begin{figure}
\centering
\includegraphics[width=1.0\linewidth]{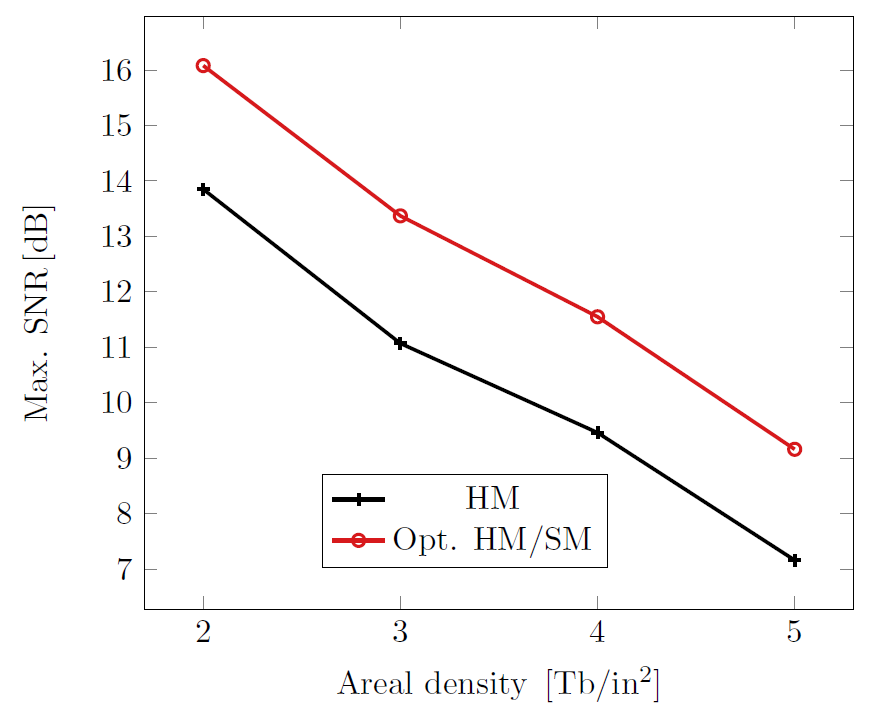}
  \caption{Maximum SNR for different areal densities for pure hard magnetic material and the optimized bilayer structure.}
  \label{opt2maxSNRoverAD}
\end{figure}

To compute the SNR for a certain $(t,b)$ combination, the reader was scaled in both the down-track and the off-track direction according to the bit length and the track width, respectively. The reader resolution $R$ in down-track direction is scaled by

\begin{align}
    R=R_0 \cdot \frac{b}{b_0}
\end{align}

where $b$ is the bit length, $R_0=13.26\,$nm is the initial reader resolution and $b_0=10.2\,$nm denotes the mean initial bit length according to ASTC. In off-track direction, the reader width is scaled to the respective track width $t$. The initial track width is $44.34\,$nm. In \Cref{opt2SNRphase}(a) and (b) the SNR is shown as a function of the bit length and the track width for pure hard magnetic material and the optimized bilayer structure, respectively. Additionally, the phase plots include the SNR curves for $(t,b)$ combinations that yield areal densities from 2 to 5\,Tb/in$^2$. From the phase diagram it is visible that higher SNR values can be achieved for the optimized structures than for the pure hard magnetic material in the same bit length $-$ track width range. For example, the SNR for an areal density of 2\,Tb/in$^2$ for the bilayer structure is larger than 15\,dB whereas it is between 10\,dB and 15\,dB for pure hard magnetic material. 
For each AD there is a $(t,b)$ combination for which the SNR is maximal and which is marked by a dot in the phase plot. In \Cref{opt2maxSNRoverAD} the maximum SNR over the areal density is displayed for both structures. The results show that the SNR that can be achieved with the optimized structure is around 2 db higher than that of the hard magnetic material, if the same areal density is assumed. 
To get the same SNR, the optimized design allows for an areal density that is 1\,Tb/in$^2$ higher than for the hard magnetic one. Summarizing, the bit length $-$ track width combinations at which the maximum SNR is achieved are given in \Cref{opt2tablemaxSNR}.

\begin{center}
\begin{table*}
\centering
\begin{tabular}{|c|c|c|c||c|c|c|c|}\hline
AD\,[Tb/in$^2$]& Max. SNR\,[dB] (HM) & $x$\,[nm] (HM)&$y$\,[nm]\,(HM) & Max. SNR\,[dB] (HM/SM)&$x$\,[nm] (HM/SM)& $y$\,[nm]\,(HM/SM)\\
\hline
2& 13.85& 10.0 & 32.26&16.08&8.06&40.02\\
  3& 11.07& 6.23& 34.52 &13.37&5.37&37.53\\
   4&9.46& 5.0 & 32.26&11.55&5.0&32.26\\
   5&7.16& 4.3& 30.01&9.16&4.69&27.51\\
    \hline
 \end{tabular}
\caption{Resulting bit length $x$ and track width $y$ combinations for the maximum SNR at different areal densities (AD) for pure hard magnetic material (HM) and the optimized bilayer structure (HM/SM).}
\label{opt2tablemaxSNR}
\end{table*}
\end{center}


\section{Conclusion}

To conclude, we optimized a recording medium with high/low $T_{\mathrm{C}}$ grains for heat-assisted magnetic recording with a low Gilbert damping in the hard magnetic part $\alpha_{\mathrm{HM}}=0.02$. The simulations for a cylindrical recording grain with $d=5$\,nm and $h=8$\,nm were performed with the atomistic simulation program VAMPIRE. The damping constant of the soft magnetic material was assumed to be enhanced by doping the soft magnetic material with transition metals. The simulations showed that larger damping constants lead to smaller jitter and higher switching probabilities. A damping constant $\alpha_{\mathrm{SM}}=0.1$, in combination with an exchange interaction $J_{ij,\mathrm{SM}}=7.72\times10^{-21}\,$J/link and an uniaxial anisotropy constant $k_{\mathrm{u,SM}}=0\,$J/link, led to the best results in terms of small down-track jitter and high switching probability in a wide range of down-track and off-track positions. Interestingly, the soft magnetic composition is almost the same as for the structure with $\alpha_{\mathrm{HM}}=0.1$ obtained in a previous work \cite{noisehamr}.\\
In further simulations the amount of hard and soft magnetic material was varied. Surprisingly, the results showed that a higher amount of soft magnetic material leads to smaller down-track jitter. This is not as expected since for $\alpha_{\mathrm{HM}}=0.1$ an increase of the soft magnetic material led to larger AC noise \cite{noisehamr}. However, it can be easily explained why a higher amount of soft magnetic material leads to better jitter results. Studying the influence of the damping constant on the down-track jitter shows that an increase of the damping constant from $0.02$ to $0.1$ reduces the down-track jitter by almost $30\%$. Additionally,the maximum switching probability increases to $1$.
Since it can be seen that higher damping leads to smaller jitter and higher maximum switching probability, it is reasonable that a higher amount of soft magnetic material with $\alpha_{\mathrm{SM}}=0.1$ leads to a better recording performance. In the former work the improved performance due to higher damping was not an issue since the damping constant was 0.1 in both layers. This explains the different ratios of hard and soft magnetic material.\\
Furthermore, we analyzed the increase of the areal density can be improved if the optimized bilayer structure is used instead of pure hard magnetic recording material. This was done by analyzing the signal-to-noise ratio (SNR). The results showed that the areal density of the optimized bilayer structure could be increased by 1\,Tb/in$^2$ to achieve the same SNR as for the pure hard magnetic structure. In other words, that means that at a certain areal density, the SNR was increased by 2\,dB by using the optimized structure. 
Concluding, the optimized bilayer structure is a promising design to increase the areal storage density by just modifying the recording material.

\section{ACKNOWLEDGEMENTS}
The authors would like to thank the Vienna Science and Technology Fund (WWTF) under grant No. MA14-044, the Advanced Storage Technology Consortium (ASTC), and the Austrian Science Fund (FWF) under grant No. I2214-N20 for financial support. The computational results presented have been achieved using the Vienna Scientific Cluster (VSC).


\begin{thebibliography}{10}

\bibitem{kobayashi}
Hiroshi Kobayashi, Motoharu Tanaka, Hajime Machida, Takashi Yano, and Uee~Myong
  Hwang.
\newblock {\em Thermomagnetic recording}.
\newblock Google Patents, August 1984.

\bibitem{mee}
C.~Mee and G.~Fan.
\newblock A proposed beam-addressable memory.
\newblock {\em IEEE Transactions on Magnetics}, 3(1):72--76, 1967.

\bibitem{rottmayer}
Robert~E. Rottmayer, Sharat Batra, Dorothea Buechel, William~A. Challener,
  Julius Hohlfeld, Yukiko Kubota, Lei Li, Bin Lu, Christophe Mihalcea, Keith
  Mountfield, and {others}.
\newblock Heat-assisted magnetic recording.
\newblock {\em IEEE Transactions on Magnetics}, 42(10):2417--2421, 2006.

\bibitem{kryder}
Mark~H. Kryder, Edward~C. Gage, Terry~W. McDaniel, William~A. Challener,
  Robert~E. Rottmayer, Ganping Ju, Yiao-Tee Hsia, and M.~Fatih Erden.
\newblock Heat assisted magnetic recording.
\newblock {\em Proceedings of the IEEE}, 96(11):1810--1835, 2008.

\bibitem{rausch_heat_2015}
Tim Rausch, Ed~Gage, and John Dykes.
\newblock Heat {Assisted} {Magnetic} {Recording}.
\newblock In Jean-Yves Bigot, Wolfgang Hübner, Theo Rasing, and Roy Chantrell,
  editors, {\em Ultrafast {Magnetism} {I}}, Springer {Proceedings} in
  {Physics}, pages 200--202. Springer International Publishing, 2015.

\bibitem{thermomagnetic}
G.~W. Lewicki and {others}.
\newblock {\em Thermomagnetic recording and magneto-optic playback system}.
\newblock Google Patents, December 1971.

\bibitem{burns}
L.~Burns Jr~Leslie and {others}.
\newblock {\em Magnetic recording system}.
\newblock Google Patents, December 1959.

\bibitem{evans2}
R.~F.~L. Evans, Roy~W. Chantrell, Ulrich Nowak, Andreas Lyberatos, and H.-J.
  Richter.
\newblock Thermally induced error: {Density} limit for magnetic data storage.
\newblock {\em Applied Physics Letters}, 100(10):102402, 2012.

\bibitem{basic}
Christoph Vogler, Claas Abert, Florian Bruckner, Dieter Suess, and Dirk
  Praetorius.
\newblock Basic noise mechanisms of heat-assisted-magnetic recording.
\newblock {\em Journal of Applied Physics}, 120(15):153901, 2016.

\bibitem{bitpatterned1}
H.~J. Richter, A.~Y. Dobin, R.~T. Lynch, D.~Weller, R.~M. Brockie, O.~Heinonen,
  K.~Z. Gao, J.~Xue, R.~J. M. v.~d. Veerdonk, P.~Asselin, and M.~F. Erden.
\newblock Recording potential of bit-patterned media.
\newblock {\em Applied Physics Letters}, 88(22):222512, May 2006.

\bibitem{bitpatterned2}
H.~J. Richter, A.~Y. Dobin, O.~Heinonen, K.~Z. Gao, R.~J. M. v~d Veerdonk,
  R.~T. Lynch, J.~Xue, D.~Weller, P.~Asselin, M.~F. Erden, and R.~M. Brockie.
\newblock Recording on {Bit}-{Patterned} {Media} at {Densities} of 1
  {Tb}/in$^2$ and {Beyond}.
\newblock {\em IEEE Transactions on Magnetics}, 42(10):2255--2260, October
  2006.

\bibitem{yen_bit_2013}
Bing~K. Yen, Jim Hennessey, Eric Freeman, Kim~Yang Lee, David~S. Kuo, and Mark
  Ostrowski.
\newblock Bit patterned media, August 2013.

\bibitem{fundamental}
Dieter Suess, Christoph Vogler, Claas Abert, Florian Bruckner, Roman Windl,
  Leoni Breth, and J.~Fidler.
\newblock Fundamental limits in heat-assisted magnetic recording and methods to
  overcome it with exchange spring structures.
\newblock {\em Journal of Applied Physics}, 117(16):163913, 2015.

\bibitem{suess}
Dieter Suess.
\newblock Micromagnetics of exchange spring media: {Optimization} and limits.
\newblock {\em Journal of magnetism and magnetic materials}, 308(2):183--197,
  2007.

\bibitem{suessexchange}
Dieter Suess, Thomas Schrefl, S.~Fähler, Markus Kirschner, Gino Hrkac, Florian
  Dorfbauer, and Josef Fidler.
\newblock Exchange spring media for perpendicular recording.
\newblock {\em Applied Physics Letters}, 87(1):012504, 2005.

\bibitem{victora}
R.~H. Victora and X.~Shen.
\newblock Exchange coupled composite media for perpendicular magnetic
  recording.
\newblock {\em IEEE Transactions on Magnetics}, 41(10):2828--2833, October
  2005.

\bibitem{wangexchange}
Jian-Ping Wang, Weikang Shen, and Jianmin Bai.
\newblock Exchange coupled composite media for perpendicular magnetic
  recording.
\newblock {\em IEEE transactions on magnetics}, 41(10):3181--3186, 2005.

\bibitem{coffey}
Kevin~Robert Coffey, Jan-Ulrich Thiele, and Dieter~Klaus Weller.
\newblock {\em ‘{Thermal} spring’magnetic recording media for writing using
  magnetic and thermal gradients}.
\newblock Google Patents, April 2005.

\bibitem{suess1}
Dieter Suess and Thomas Schrefl.
\newblock Breaking the thermally induced write error in heat assisted recording
  by using low and high {Tc} materials.
\newblock {\em Applied Physics Letters}, 102(16):162405, 2013.

\bibitem{areal}
Christoph Vogler, Claas Abert, Florian Bruckner, Dieter Suess, and Dirk
  Praetorius.
\newblock Areal density optimizations for heat-assisted magnetic recording of
  high-density media.
\newblock {\em Journal of Applied Physics}, 119(22):223903, 2016.

\bibitem{noisehamr}
O.~Muthsam, C.~Vogler, and D.~Suess.
\newblock Noise reduction in heat-assisted magnetic recording of bit-patterned
  media by optimizing a high/low {Tc} bilayer structure.
\newblock {\em Journal of Applied Physics}, 122(21):213903, 2017.

\bibitem{ASTC}
{ASTC} {\textbar} {IDEMA}.
\newblock \url{http://idema.org/?cat=10}.

\bibitem{evans}
Richard~FL Evans, Weijia~J. Fan, Phanwadee Chureemart, Thomas~A. Ostler,
  Matthew~OA Ellis, and Roy~W. Chantrell.
\newblock Atomistic spin model simulations of magnetic nanomaterials.
\newblock {\em Journal of Physics: Condensed Matter}, 26(10):103202, 2014.

\bibitem{mryasov2005temperature}
Oleg~N Mryasov, Ulrich Nowak, K~Yu Guslienko, and Roy~W Chantrell.
\newblock Temperature-dependent magnetic properties of fept: Effective spin
  hamiltonian model.
\newblock {\em EPL (Europhysics Letters)}, 69(5):805, 2005.

\bibitem{hovorka2012curie}
O~Hovorka, S~Devos, Q~Coopman, WJ~Fan, CJ~Aas, RFL Evans, Xi~Chen, G~Ju, and
  RW~Chantrell.
\newblock The curie temperature distribution of fept granular magnetic
  recording media.
\newblock {\em Applied Physics Letters}, 101(5):052406, 2012.

\bibitem{doping}
W.~Zhang, S.~Jiang, P.~K.~J. Wong, L.~Sun, Y.~K. Wang, K.~Wang, M.~P. de~Jong,
  W.~G. van~der Wiel, G.~van~der Laan, and Y.~Zhai.
\newblock Engineering {Gilbert} damping by dilute {Gd} doping in soft magnetic
  {Fe} thin films.
\newblock {\em Journal of Applied Physics}, 115(17):17A308, May 2014.

\bibitem{ingvarsson_tunable_2004}
S.~Ingvarsson, Gang Xiao, S.~S.~P. Parkin, and R.~H. Koch.
\newblock Tunable magnetization damping in transition metal ternary alloys.
\newblock {\em Applied Physics Letters}, 85(21):4995--4997, November 2004.

\bibitem{fassbender_structural_2006}
J.~Fassbender, J.~von Borany, A.~Mücklich, K.~Potzger, W.~Möller, J.~McCord,
  L.~Schultz, and R.~Mattheis.
\newblock Structural and magnetic modifications of {Cr}-implanted {Permalloy}.
\newblock {\em Physical Review B}, 73(18), May 2006.

\bibitem{bailey_control_2001}
W.~Bailey, P.~Kabos, F.~Mancoff, and S.~Russek.
\newblock Control of magnetization dynamics in {Ni}/sub 81/{Fe}/sub 19/ thin
  films through the use of rare-earth dopants.
\newblock {\em IEEE Transactions on Magnetics}, 37(4):1749--1754, July 2001.

\bibitem{rantschler_effect_2007}
J.~O. Rantschler, R.~D. McMichael, A.~Castillo, A.~J. Shapiro, W.~F. Egelhoff,
  B.~B. Maranville, D.~Pulugurtha, A.~P. Chen, and L.~M. Connors.
\newblock Effect of 3d, 4d, and 5d transition metal doping on damping in
  permalloy thin films.
\newblock {\em Journal of Applied Physics}, 101(3):033911, February 2007.

\bibitem{wang_transition_2009}
Xiaobin Wang, Bogdan Valcu, and Nan-Hsiung Yeh.
\newblock Transition width limit in magnetic recording.
\newblock {\em Applied Physics Letters}, 94(20):202508, 2009.

\bibitem{varvaro_ultra-high-density_2016}
Gaspare Varvaro and Francesca Casoli.
\newblock {\em Ultra-{High}-{Density} {Magnetic} {Recording}: {Storage}
  {Materials} and {Media} {Designs}}.
\newblock CRC Press, March 2016.

\bibitem{slanovc}
Florian Slanovc, Christoph Vogler, Olivia Muthsam, and Dieter Suess.
\newblock Systematic parameterization of heat-assisted magnetic recording
  switching probabilities and the consequences for the resulting snr.
\newblock {\em arXiv preprint arXiv:1907.03884}, 2019.

\bibitem{hagedorn_analysis_1970}
F.~B. Hagedorn.
\newblock Analysis of {Exchange}‐{Coupled} {Magnetic} {Thin} {Films}.
\newblock {\em Journal of Applied Physics}, 41(6):2491--2502, May 1970.

\bibitem{suess_multilayer_2006}
D.~Suess.
\newblock Multilayer exchange spring media for magnetic recording.
\newblock {\em Applied Physics Letters}, 89(11):113105, September 2006.

\bibitem{vogler_landau-lifshitz-bloch_2014}
Christoph Vogler, Claas Abert, Florian Bruckner, and Dieter Suess.
\newblock Landau-{Lifshitz}-{Bloch} equation for exchange-coupled grains.
\newblock {\em Physical Review B}, 90(21):214431, December 2014.

\bibitem{hernandez_using_2017}
S.~Hernández, P.~Lu, S.~Granz, P.~Krivosik, P.~Huang, W.~Eppler, T.~Rausch,
  and E.~Gage.
\newblock Using {Ensemble} {Waveform} {Analysis} to {Compare} {Heat} {Assisted}
  {Magnetic} {Recording} {Characteristics} of {Modeled} and {Measured}
  {Signals}.
\newblock {\em IEEE Transactions on Magnetics}, 53(2):1--6, February 2017.

\end{thebibliography}

\end{document}